\documentclass[a4paper]{article}
\usepackage[utf8]{inputenc}
\usepackage{pgfplots}
\usepackage{tikz}
\usepackage{acronym}
\usepackage{amsmath}
\usepackage{amssymb}
\usepackage{booktabs}
\usepackage{bigfoot}
\usepackage{cprotect}
\usepackage{cancel}
\usepackage{subcaption}
\usepackage{url}
\usepackage{hyperref}
\usepackage[capitalise]{cleveref}

\title{Zero-Liquidation Loans:\\A Structured Product Approach to DeFi Lending}
\author{Aetienne Sardon\\\small{as@myso.finance}}

\date{October 2021\\\small{V1}}

\begin{document}

\maketitle

\newacro{AMM}[AMM]{Automated Market Maker}
\newcommand{\AMM}{\ac{AMM}\xspace}
\newcommand{\AMMs}{\acp{AMM}\xspace}

\newacro{DeFi}[DeFi]{Decentralized Finance}
\newcommand{\DEFI}{\ac{DeFi}\xspace}

\newacro{DEX}[DEX]{Decentralied Exchange}
\newcommand{\DEX}{\ac{DEX}\xspace}
\newcommand{\DEXs}{\acp{DEX}\xspace}

\newacro{EWMA}[EWMA]{Exponential Weighted Moving Average}
\newcommand{\EWMA}{\ac{EWMA}\xspace}

\newacro{IRS}[IRS]{Interest Rate Swap}
\newcommand{\IRS}{\ac{IRS}\xspace}
\newcommand{\IRSs}{\acp{IRS}\xspace}

\newacro{LTV}[LTV]{Loan to Value}
\newcommand{\LTV}{\ac{LTV}\xspace}

\newacro{LP}[LP]{Liquidity Pool}
\acrodefplural{LPs}{Liquidity Pools}
\newcommand{\LP}{\ac{LP}\xspace}
\newcommand{\LPs}{\acp{LP}\xspace}

\newacro{LS}[LS]{Liqiudity Staker}
\newcommand{\LS}{\ac{LS}\xspace}
\newcommand{\LSs}{\acp{LS}\xspace}

\newacro{PnL}[PnL]{Profit and Loss}
\newcommand{\PnL}{\ac{PnL}\xspace}

\newacro{SC}[SC]{Smart Contract}
\newcommand{\SC}{\ac{SC}\xspace}

\newacro{TVL}[TVL]{Total Value Locked}
\newcommand{\TVL}{\ac{TVL}\xspace}

\newacro{TradFi}[TradFi]{Traditional Finance}
\newcommand{\TradFi}{\ac{TradFi}\xspace}

\newacro{VaR}[VaR]{Value at Risk}
\newcommand{\VAR}{\ac{VaR}\xspace}

\newacro{YSP}[YSP]{Yield Swap Protocol}
\newcommand{\YSP}{\ac{YSP}\xspace}

\newcommand{\COMMA}{\,,\xspace}
\newcommand{\PERIOD}{\,.}

\newcommand{\ATOKEN}{{\tt{aToken}}\xspace}

\newcommand{\TODO}[1]{\colorbox{yellow}{#1}}
\newcommand{\TOFIX}[1]{\colorbox{pink}{#1}}

\begin{abstract}
Zero-liquidation loans allow users to borrow \verb|USDC| against their \verb|ETH| holdings, but without the risk of being liquidated in case of LTV shortfalls. This is achieved by giving users the option to repay their loans, either in \verb|USDC| or through their previously pledged \verb|ETH| (the concept can be generalized to other currency pairs as well). Liquidity providers, on the other hand side, are compensated with a higher yield for bearing the \verb|ETH| downside risk. A positive side effect of zero-liquidation loans is that they are more robust against flash crashes and have a lower risk of triggering financial contagion effects than currently prevailing liquidation-centered design approaches for lending and borrowing in DeFi. 
\end{abstract}

\section{Introduction}
The demand for crypto-collateralized lending and borrowing in DeFi has surged, recently surpassing a TVL of \$40bn in 2021 \cite{aave} \cite{compound}. However, with currently available borrowing products borrowers face two pain points: (i) liquidations and (ii) variability in rates. While some solutions already exist to alleviate (ii), there currently aren't any products to spare users from (ii). Zero-liquidation loans solve both.\\

On a YTD basis there have been 9,023 liquidations on Aave with a 7-day liquidation amount of approx. \$37mn \cite{dune}. Liquidations occur when the collateral that borrowers pledge suddenly drops in value. Borrowers, who fail to prevent LTV shortfalls face the risk of having their collateral liquidated and being penalized with liquidation fees. Moreover, commonly used incentive mechanisms tend to favor liquidators over borrowers, causing the problem of so called over-liquidation, leading to unnecessary high losses for borrowers \cite{qin}. In order to avoid liquidations, borrowers need to constantly monitor their LTV and remain alert to quickly respond to changing market conditions. When there are many users with a leveraged long position in the collateral currency, even a random dip in market prices can cause a whole cascade of liquidations, leading to a self-accelerating selling pressure. Such market situations may lead to network congestion and hiking gas costs, as has been the case in the \verb|ETH| market collapse of March 13th 2020 \cite{qin}, leaving some borrowers unable to react despite imminent liquidations. For borrowers who get liquidated it can be particularly annoying if market prices recover after a dip again, leaving them deprived from subsequent upward price participation.\\


\section{Zero-Liquidation Loans}
\label{sec:zero_liquidation_loans}
Zero-liquidation loans resemble so called \emph{double currency notes}, which give borrowers the option to repay a loan in either one of two currencies (e.g., \verb|ETH| or \verb|USDC|). This allows users to borrow funds against their crypto holdings, but without being exposed to any liquidation risk. Liquidity providers, on the other hand side, are compensated with a higher yield.\\

\begin{figure}
    \centering
    \includegraphics[width=1.0\textwidth]{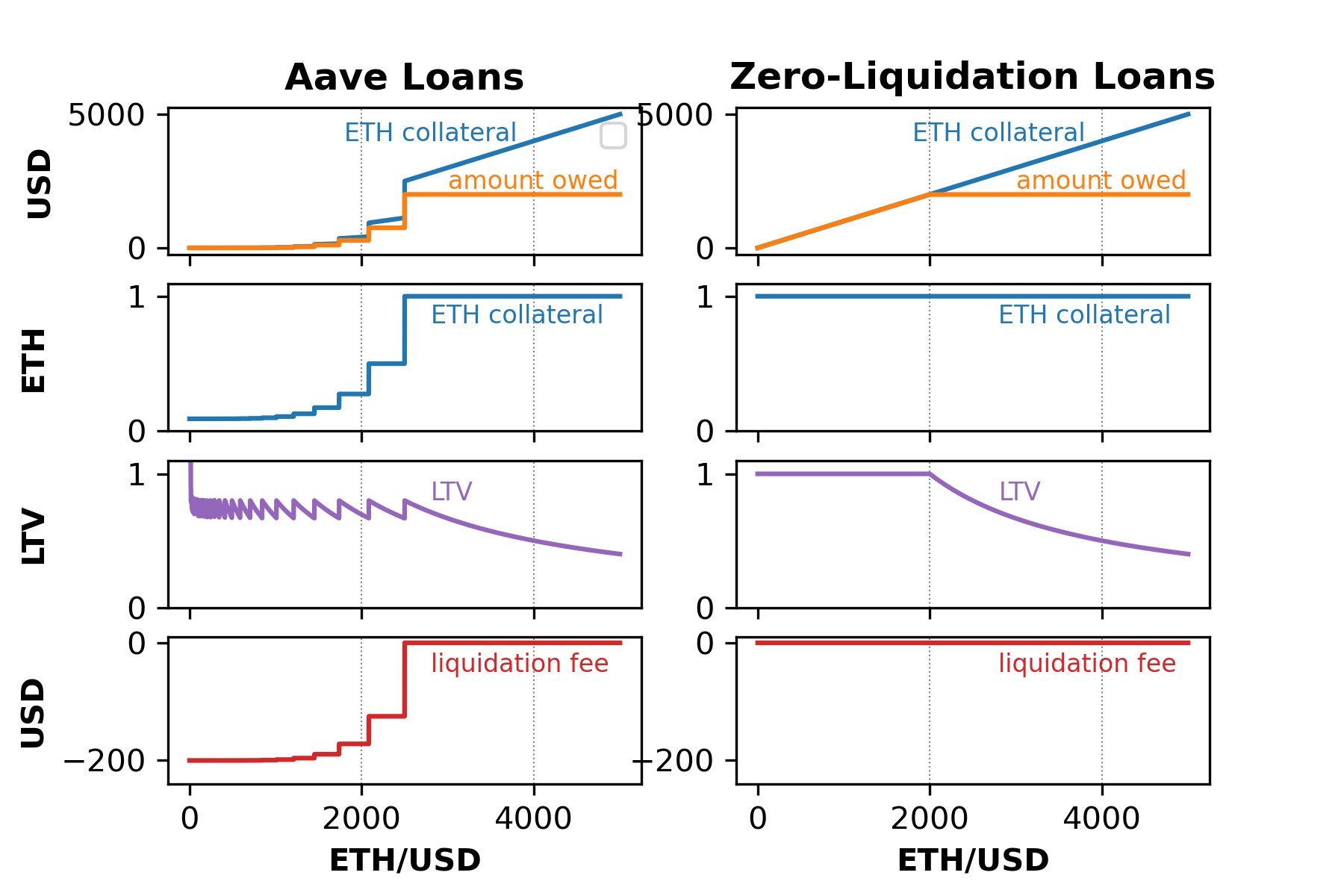} 
    \cprotect\caption{\small Stylized comparison of an Aave vs. a zero-liquidation loan, given an initial loan amount of \verb|2,000 USDC| against \verb|1 ETH| in collateral at a spot price of \verb|4,000 ETH/USDC|, with a LTV threshold of 80\% and liquidation penalty fee of 5\%.}
    \label{fig:aave_vs_zero_liquidation}
\end{figure}

Zero-liquidation loans stand in contrast to currently existing lending and borrowing solutions in DeFi, where users need to constantly monitor their LTV, and hedge against potential shortfalls to avoid having to pay liquidation penalty fees \cite{willtoshower} \cite{interoperate}. \cref{fig:aave_vs_zero_liquidation} illustrates the differences between an Aave and a zero-liquidation loan. It can be seen that borrowers on Aave will have their \verb|ETH| collateral liquidated every time the LTV reaches the liquidation threshold (e.g., 80\%). As a result, an Aave borrower's \verb|ETH| holding will be reduced if the \verb|ETH| price falls to a certain threshold, even if the price moves up again afterwards. In addition, borrowers will be charged a fee every time a liquidation happens. In contrast, with zero-liquidation loans the borrowers' \verb|ETH| holdings remain constant and they don't incur any liquidation fees. At the same time, borrowers still have full upside and only owe the initially borrowed amount or the pledged collateral, whichever is lower.

\subsection{Borrower's Perspective}
\label{sec:borrower_perspective}
Assume Bob wants to take out a 90 day loan against his crypto holdings of \verb|1 ETH|, where \verb|1 ETH| is currently worth \verb|4,000 USDC|. With a zero-liquidation loan, he pledges \verb|1 ETH| and receives, e.g., \verb|1,850 USDC| against it as a loan (the exact amount will depend on supply and demand). Then, after 90 days he has the option to reclaim his collateral for a pre-agreed repayment amount of, e.g., \verb|2,000 USDC|. The repayment amount includes the interest cost and is thus higher than the amount previously paid out. So Bob effectively has fixed borrowing costs of \verb|150 USDC|, with an implied APR of 30\%. However, in contrast to existing DeFi lending solutions Bob isn't exposed to changing borrowing costs and, more importantly, is spared from liquidations. This is because Bob can later choose whether he wants to repay his loan in \verb|USDC| or \verb|ETH|.\\

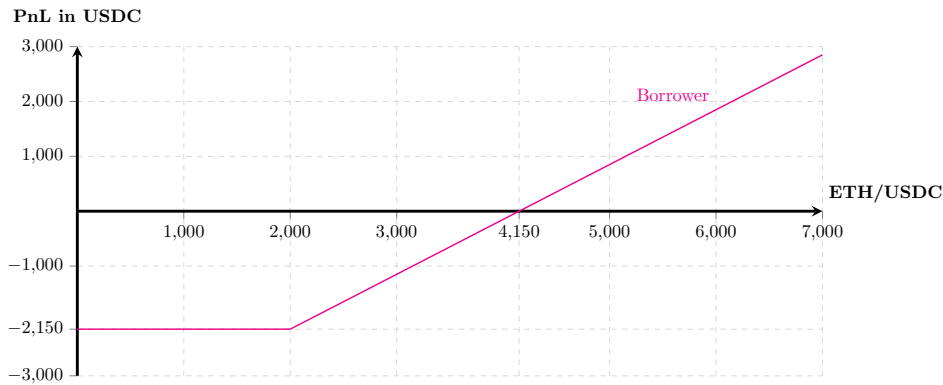
\begin{figure}
\begin{tikzpicture}[scale = 0.68]
    \begin{axis}[
        axis x line=middle,
        axis y line=middle,
        axis line style = ultra thick,
        grid = major,
        width=16cm,
        height=8cm,
        grid style={dashed, gray!30},
        xmin=0,     
        xmax=7000,    
        ymin=-3000,     
        ymax=3000,   
        xlabel style={at={(1,0.48)},above right,yshift=5pt},
        ylabel style={at={(0,1)},above,yshift=10pt},
        xticklabel style={
                /pgf/number format/fixed,
                /pgf/number format/precision=2
        },
        scaled x ticks=false,
        yticklabel style={
                /pgf/number format/fixed,
                /pgf/number format/precision=2
        },
        scaled y ticks=false,
        xlabel=\textbf{ETH/USDC},
        ylabel=\textbf{PnL in USDC},
		/pgfplots/xtick={0, 1000, 2000, 3000, 4150, 5000, 6000, 7000}, 
		/pgfplots/ytick={-3000, -2150, -1000, 0, 1000, 2000, 3000}, 
        tick align=outside,
        enlargelimits=false]
      \addplot[domain=0:2000, magenta, thick,samples=100] {-150-2000};
      \addplot[domain=2000:7000, magenta, thick,samples=100] {(x-2000)-2150} node[left,pos=0.8,yshift=8pt] {Borrower};
	  \end{axis}
\end{tikzpicture}
\caption{\small Bob's loan PnL after 90 days, assuming a loan amount of 2,000 USDC, secured with 1 ETH (worth 4,000 USDC at inception), and interest costs of 150 USDC.}
\label{fig:zero_liquidation_loan}
\end{figure}

More specifically, after 90 days Bob can either pay back \verb|2,000 USDC| and receive his originally pledged \verb|1 ETH| back, or, alternatively, simply walk away from the loan. Naturally, Bob will chose whatever option is better for him, i.e., if the price of \verb|1 ETH| is higher than \verb|2,000 USDC|, then he will chose to repay the loan in \verb|USDC|, whereas, if the price of \verb|1 ETH| is below \verb|2,000 USDC| he will be better off leaving his \verb|1 ETH| collateral to the lender, which effectively means he's repaying in \verb|ETH|. Bob's resulting payoff is illustrated in \cref{fig:zero_liquidation_loan}. It can be seen that, even if the \verb|ETH| price drops below \verb|2,000 USDC|, his downside is capped because by having borrowed \verb|1,850 USDC| he has reduced his \verb|ETH| exposure to \verb|4,000 USDC - 1,850 USDC = 2,150 USDC|.\\

Note that while Bob isn't exposed to any liquidation penalty fees, he still has full \verb|ETH| upside. Bob obviously can use the borrowed \verb|1,850 USDC| for whatever he wants, e.g., yield farming or also leveraging his \verb|ETH| position by repeating the previously described borrowing procedure, analogous to how users do it today on lending and borrowing platforms like Aave etc. 

\subsection{Liquidity Provider's Perspective}
Assume Larry holds \verb|2,000 USDC| and wants to boost his APY. He provides his \verb|USDC| to the zero-liquidation loan pool and receives a corresponding pro-rata share of the pool's PnL. For example, assume the pool currently has a TVL of \verb|4,000 USDC|, then Larry's pledged liquidity corresponds to a 50\% share. If now someone borrows, e.g., \verb|2,000 USDC| for 90 days from this pool at a borrowing cost of \verb|150 USDC| (see example in \cref{sec:borrower_perspective}), then Larry will be entitled to a 50\% pro-rata share of the thereof resulting PnL.\\

\begin{figure}
\begin{tikzpicture}[scale = 0.68]
    \begin{axis}[
        axis x line=middle,
        axis y line=middle,
        axis line style = ultra thick,
        grid = major,
        width=16cm,
        height=8cm,
        grid style={dashed, gray!30},
        xmin=0,     
        xmax=7000,    
        ymin=-1000,     
        ymax=200,   
        xlabel style={at={(1,0.78)},above right,yshift=5pt},
        ylabel style={at={(0,1)},above,yshift=10pt},
        xticklabel style={
                /pgf/number format/fixed,
                /pgf/number format/precision=2
        },
        scaled x ticks=false,
        yticklabel style={
                /pgf/number format/fixed,
                /pgf/number format/precision=2
        },
        scaled y ticks=false,
        xlabel=\textbf{ETH/USDC},
        ylabel=\textbf{PnL in USDC},
		/pgfplots/xtick={0, 1000, ..., 7000}, 
		/pgfplots/ytick={-1000, -925, -500, 0, 75, 200}, 
        tick align=outside,
        enlargelimits=false]
      \addplot[domain=0:2000, blue, thick,samples=100] {-1000+x*0.5+75};
      \addplot[domain=2000:7000, blue, thick,samples=100] {75} node[above,pos=0.58] {Liquidity Provider};
	  \end{axis}
\end{tikzpicture}
\caption{\small Larry's PnL after 90 days, assuming interest earnings of 150 USDC and a pool share of 50\%.}
\label{fig:zero_liquidation_loan_deposit}
\end{figure}
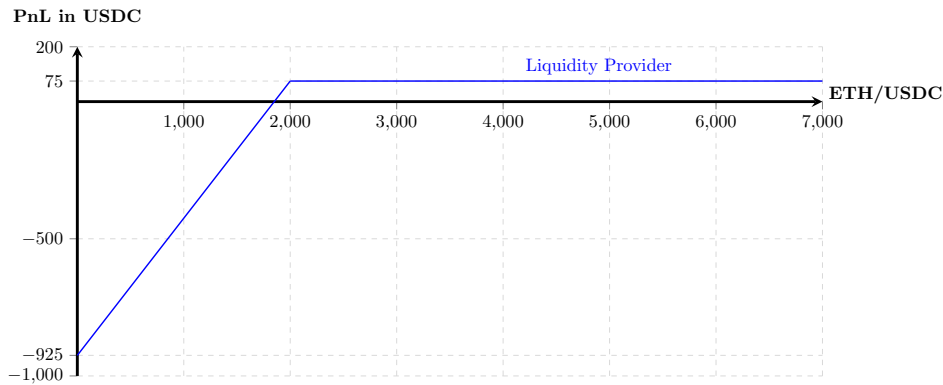

More specifically, if after 90 days the \verb|ETH| price is above \verb|2,000 USDC| then the pool will have earned \verb|150 USDC|, in which case Larry will be credited \verb|75 USDC| (i.e., his 50\% share), implying a 15\% APY. Only if the \verb|ETH| price drops by more than 50\% (the initial LTV), i.e., below \verb|2,000 USDC|, Larry will suffer a loss, which, however, is capped at \verb|-925 USDC|, i.e., even in the most adverse scenario where the \verb|ETH| price drops to zero. \cref{fig:zero_liquidation_loan_deposit} shows the resulting payoff.

\section{Option Representation}
Another way to think about crypto-collateralized loans is to view them as options. \cref{sec:zero_liquididation_loans_and_vanilla_options} and \cref{sec:aave_loans_and_barrier_options} describe how users can synthetically replicate crypto-collateralized loan payoffs using different types of options. Note, however, that these replication strategies shall only serve as a mental model to help later construct the AMM (see \cref{sec:AMM}), but may have shortcomings when used in practice.\footnote{For example, for users who want to physically hold the underlying such replication strategies create transactional overhead and costs. I.e., a user would have to sell the collateral, buy a corresponding option, and, eventually, buy back the collateral in case the option is cash-settled.}

\subsection{Zero-Liquidation Loans and Vanilla Options}
\label{sec:zero_liquididation_loans_and_vanilla_options}
A user borrowing through a zero-liquidation loan is equivalent to a user that initially holds a crypto asset, then sells it in the spot market, buys an option on it and keeps the difference between the underlying sale proceeds and the option purchase price as a short term funding source. In other words, a zero-liquidation loan borrower can be represented as someone holding some underlying that he wants to swap into a call option and cash amount, i.e.
\begin{equation}
\small
\begin{split}
Loan &= C_K+K
\end{split}
\end{equation}
where $C_K$ denotes the call option with strike price $K$ and where $K$ is also equal to borrowed cash amount. This portfolio allows the borrower to maintain his upside on the collateral (=call option), while at the same time receiving a cash amount against it (=strike price).\\ 

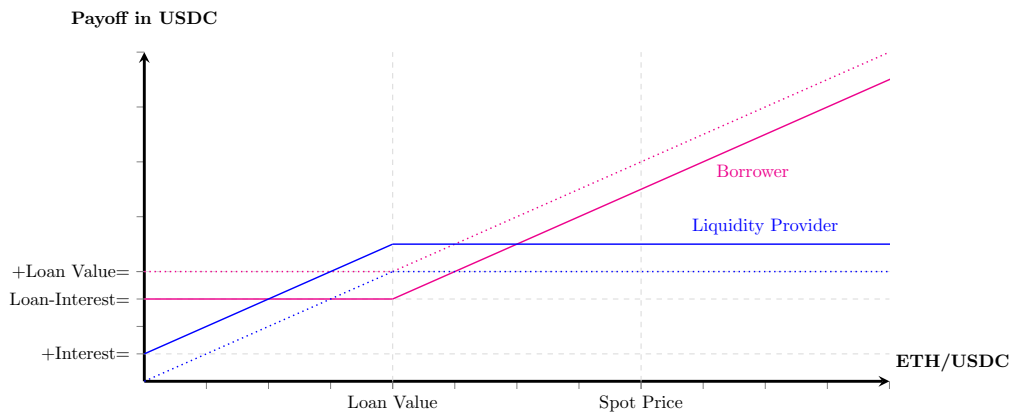
\begin{figure}
\begin{tikzpicture}[scale = 0.68]
    \begin{axis}[
        axis x line=middle,
        axis y line=middle,
        axis line style = ultra thick,
        width=16cm,
        height=8cm,
        grid style={dashed, gray!30},
        xmin=0,     
        xmax=6000,    
        ymin=0,     
        ymax=6000,   
        xlabel style={at={(1,0)},above right,yshift=2pt},
        ylabel style={at={(0,1)},above,yshift=10pt},
        xticklabel=\empty,
        scaled x ticks=false,
        yticklabel=\empty,
        scaled y ticks=false,
        xlabel=\textbf{ETH/USDC},
        ylabel=\textbf{Payoff in USDC},
		/pgfplots/xtick={},
		/pgfplots/ytick={}, 
        tick align=outside,
        ,extra y ticks={500, 1500, 2000}
        ,extra y tick style={%
            ,grid=major
            ,ticklabel pos=top}
        ,extra y tick labels={+Interest=, Loan-Interest=, +Loan Value=},
        ,extra x ticks={2000, 4000}
        ,extra x tick style={%
            ,grid=major
            ,ticklabel pos=top}
        ,extra x tick labels={Loan Value, Spot Price},
        enlargelimits=false]
      \addplot[domain=0:2000, magenta, thick,dotted,samples=100] {2000};
      \addplot[domain=2000:7000, magenta, thick,dotted,samples=100] {2000+(x-2000)} node[below,pos=0.58,yshift=-25pt] {Borrower};
      \addplot[domain=0:2000, magenta, thick,samples=100] {2000-500};
      \addplot[domain=2000:7000, magenta, thick,samples=100] {2000+(x-2000)-500};
      
      \addplot[domain=0:2000, blue, thick,dotted,samples=100] {x};
      \addplot[domain=2000:7000, blue, thick,dotted,samples=100] {2000};
      \addplot[domain=0:2000, blue, thick,samples=100] {x+500};
      \addplot[domain=2000:7000, blue, thick,samples=100] {2000+500} node[above,pos=0.6,yshift=1.5pt] {Liquidity Provider};
      
      
	  \end{axis}
\end{tikzpicture}
\caption{\small Stylized payoff diagrams for borrowers and liquidity providers in zero-liquidation loans. Note that the interest is essentially the put option premium.}
\label{fig:stylized_payoff_diagrams}
\end{figure}

\cref{fig:stylized_payoff_diagrams} illustrates the borrower's payoff. It can be seen that the borrower's downside is limited, i.e., even if the underlying's value drops to zero the net asset position is equal to at least the loan amount he was paid out by the lender. This is different to the downside borrowers face with Aave, where adverse price movements in the underlying may cause liquidations and corresponding penalty fees, and, lead to foregone upside participation if prices bounce back after breaching liquidation thresholds.\\


On the other hand side, liquidity providers of zero-liquidation loans give borrowers the option to repay either with their pledged collateral or the original loan value. Naturally, borrowers will choose whichever option is better for them. Therefore liquidity providers can expect to receive a repayment amount of
\begin{equation}
\small
\begin{split}
Repayment &= \min(S_T, K) \\
&= K-\underbrace{\max(K-S_T, 0)}_{=P_K}
\end{split}
\end{equation}
where $S_T$ denotes the collateral's price at time $T$ and $K$ represents the received loan amount (and $P_K$ a put option). The resulting payoff is shown in \cref{fig:stylized_payoff_diagrams}. 


\subsection{Aave Loans and Barrier Options}
\label{sec:aave_loans_and_barrier_options}
Not only liquidation loans can be represented through options (see \cref{sec:zero_liquididation_loans_and_vanilla_options}), but in fact, Aave loans as well. The difference, however, is that replicating Aave loans requires a portfolio of barrier options instead of vanilla options.\\

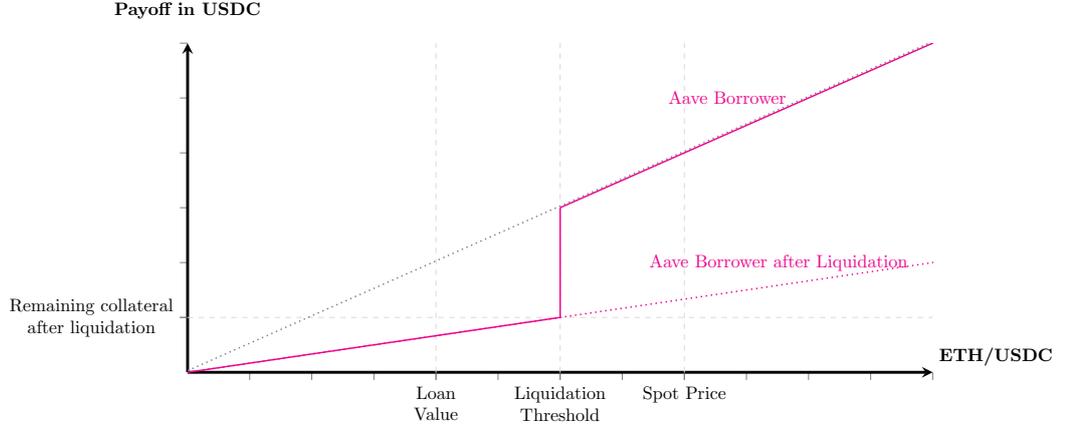
\begin{figure}
\begin{tikzpicture}[scale = 0.68]
    \begin{axis}[
        axis x line=middle,
        axis y line=middle,
        axis line style = ultra thick,
        width=16cm,
        height=8cm,
        grid style={dashed, gray!30},
        xmin=0,     
        xmax=6000,    
        ymin=0,     
        ymax=6000,   
        xlabel style={at={(1,0)},above right},
        ylabel style={at={(0,1)},above,yshift=10pt},
        xticklabel=\empty,
        scaled x ticks=false,
        yticklabel=\empty,
        scaled y ticks=false,
        xlabel=\textbf{ETH/USDC},
        ylabel=\textbf{Payoff in USDC},
		/pgfplots/xtick={},
		/pgfplots/ytick={}, 
        tick align=outside,
        ,extra y ticks={1000}
        ,extra y tick style={%
            ,grid=major
            ,ticklabel pos=top, style={align=center}}
        ,extra y tick labels={Remaining collateral \\after liquidation},
        ,extra x ticks={2000, 3000, 4000}
        ,extra x tick style={%
            ,grid=major
            ,ticklabel pos=top, style={align=center}}
        ,extra x tick labels={Loan\\Value, {Liquidation\\Threshold}, Spot Price},
        enlargelimits=false]
      \addplot[domain=0:7000, gray, thick,dotted,samples=100] {x+30};
      \addplot[domain=0:3000, magenta, thick,samples=100] {x/3} node[below,pos=0.58,yshift=-18pt] {Borrower};
      \addplot[domain=0:3000,thick,samples=50,smooth,magenta] coordinates {(3000,1000)(3000,3000)};
      \addplot[domain=3000:7000, magenta, thick,samples=100] {x} node[left,pos=0.5,xshift=-8.5pt] {Aave Borrower};
      \addplot[domain=0:7000, magenta, thick, dotted,samples=100] {x/3} node[above,pos=0.68,yshift=4.5pt] {Aave Borrower after Liquidation};
	  \end{axis}
\end{tikzpicture}
\caption{\small Stylized diagram of a simplified Aave borrower payoff with a 100\% liquidation. The payoff resembles a down-and-out call option, where the liquidation threshold corresponds to the barrier and the loan value to the strike price. After liquidation the payoff is altered because of the decreased crypto collateral position.}
\label{fig:stylized_payoff_diagram_Aave}
\end{figure}

In order to see this, let's assume there's an Aave loan that liquidates 100\% of the collateral in case the liquidation threshold is reached. The resulting payoff is shown in \cref{fig:stylized_payoff_diagram_Aave}. As long as the collateral's price is above the LTV threshold, the borrower participates 1:1 with the underlying price movement. However, in case the LTV barrier is breached, the collateral is liquidated and the user only participates with the remaining collateral. Thus, a user holding a linear combination of a down-and-out call plus a down-and-in call option would have the same payoff. In practice, however, Aave doesn't liquidate 100\% of collateral, but only 50\%. An exact payoff replication would therefore require a linear combination of several multi-barrier options, where the down-and-in and down-and-out barriers would correspond to the subsequent liquidation thresholds.\footnote{For example, in case of a LTV threshold of 80\%, one would need a down-and-out call with a barrier at LTV=80\% and a multi-barrier option with a down-and-in barrier at LTV=80\%, as well as a down-and-out barrier at LTV=64\% (=80\% of 80\%) and half the nominal, and so on.}

\section{Automated Market Making}
\label{sec:AMM}
As described in \cref{sec:zero_liquididation_loans_and_vanilla_options}, a borrower taking out a zero-liquidation loan is equivalent to a user swapping the underlying $S$ for $C_K+K$. \cref{fig:amm1} illustrates the steps involved in the swap. In the following sections we will describe (i) how to determine the borrowable amounts per provided collateral unit, and, (ii) how to determine the strike price for the embedded call option. These two results can then be combined to construct an AMM for trading zero-liquidation loans.


\begin{figure}
\begin{center}
\begin{tikzpicture}[thick,scale=0.8, every node/.style={transform shape}]

\node (Borrower) at (0,0) [rectangle,draw,rounded corners,minimum width=3.2cm,minimum height=1.8cm,outer sep=2.8] {\begin{tabular}{c}\textbf{Borrower}\\inventory:\\ $S$\end{tabular}};

\node (AMM) at (6,0) [rectangle,draw,rounded corners,minimum width=3.2cm,minimum height=1.8cm,outer sep=2.8] {\begin{tabular}{c}\textbf{AMM}\\inventory:\\ $K$\end{tabular}};

\end{tikzpicture}
\caption{Initial state: the borrower holds the underlying $S$ and the AMM has lendable liquidity $K$.}
\label{fig:amm1}

\begin{tikzpicture}[thick,scale=0.8, every node/.style={transform shape}]

\node (Borrower) at (0,0) [rectangle,draw,rounded corners,minimum width=3.2cm,minimum height=1.8cm,outer sep=2.8] {\begin{tabular}{c}\textbf{Borrower}\\inventory:\\\footnotesize$\cancel{S}\textcolor{red}{-\cancel{S}}$\\\footnotesize$\textcolor{black!30!green}{+(C_K+K)}$\end{tabular}};

\node (AMM) at (6,0) [rectangle,draw,rounded corners,minimum width=3.2cm,minimum height=1.8cm,outer sep=2.8] {\begin{tabular}{c}\textbf{AMM}\\inventory:\\ \footnotesize$\cancel{K}\textcolor{red}{-(C_K+\cancel{K})}$\\\footnotesize$\textcolor{black!30!green}{+S}$\end{tabular}};

\draw [->] (Borrower) edge[bend left] (AMM) node[xshift=86,yshift=48]{$S$};
\draw [<-] (Borrower) edge[bend right] (AMM) node[xshift=85,yshift=-48]{$C_K+K$};

\end{tikzpicture}
\caption{Borrower takes out a zero-liquidation loan, which is equivalent to swapping the underlying $S$ for a call option $C_K$ and the loan amount $K$.}
\label{fig:amm2}

\begin{tikzpicture}[thick,scale=0.8, every node/.style={transform shape}]

\node (Borrower) at (0,0) [rectangle,draw,rounded corners,minimum width=3.2cm,minimum height=2cm,outer sep=2.8] {\begin{tabular}{c}\textbf{Borrower}\\inventory:\\ \footnotesize $\textcolor{black!30!green}{+S}\textcolor{red}{-\cancel{K}}$\\\footnotesize $+(\cancelto{exercised}{C_K}+\cancel{K})$\end{tabular}};

\node (AMM) at (6,0) [rectangle,draw,rounded corners,minimum width=3.2cm,minimum height=2cm,outer sep=2.8] {\begin{tabular}{c}\textbf{AMM}\\inventory:\\ \footnotesize $\textcolor{black!30!green}{+K}+\cancel{S}\textcolor{red}{-\cancel{S}}$\end{tabular}};

\draw [->] (Borrower) edge[bend left] (AMM) node[xshift=86,yshift=48]{$K$};
\draw [<-] (Borrower) edge[bend right] (AMM) node[xshift=85,yshift=-48]{$S$};

\end{tikzpicture}
\caption{Repayment scenario $S_T>K$: the borrower repays the loan to reclaim his collateral.}
\label{fig:amm3}

\begin{tikzpicture}[thick,scale=0.8, every node/.style={transform shape}]

\node (Borrower) at (0,0) [rectangle,draw,rounded corners,minimum width=3.2cm,minimum height=1.8cm,outer sep=2.8] {\begin{tabular}{c}\textbf{Borrower}\\inventory:\\\footnotesize $(\cancelto{unexercised}{C_K}+K)$\end{tabular}};

\node (AMM) at (6,0) [rectangle,draw,rounded corners,minimum width=3.2cm,minimum height=1.8cm,outer sep=2.8] {\begin{tabular}{c}\textbf{AMM}\\inventory:\\ \footnotesize $S$\end{tabular}};

\end{tikzpicture}
\caption{Repayment scenario $S_T \leq K$: the borrower leaves the option unexercised and walks away from the collateral.}
\label{fig:amm4}
\end{center}
\end{figure}

\subsection{Determining borrowable Amounts and Swap Quantities}
\label{sec:determining_borrowable_amounts}

Let $Q_{\mathcal{C}}$ denote the quantity of crypto collateral and $Q_\mathcal{B}$ be the quantity of borrow currency one can borrow. Applying the constant product formula one can construct an AMM that swaps $Q_{\mathcal{C}}$ and $Q_\mathcal{B}$ according to
\begin{equation}
\begin{split}
(Q_\mathcal{C}+\Delta Q_\mathcal{C})(Q_\mathcal{B}-\Delta Q_\mathcal{B}) &= k
\end{split}
\end{equation}
, where $\Delta Q_\mathcal{C}$ and $\Delta Q_\mathcal{B}$ denote the corresponding quantity changes and $k$ an AMM constant. But how do these quantities relate to the quantities of options the borrower receives? Obviously, the borrower will want to have an option for every collateral unit he gives to the AMM. So if he gives $\Delta Q_{\mathcal{C}}$ units of $S$, he expects to receive the same quantity of $C_K$. Luckily, the AMM isn't per se limited on the number of options it can offer to borrowers. This is easy to see when considering that if $K=0$ the AMM can always write an option for every collateral unit $\Delta Q_{\mathcal{C}}$ it receives (this is because a call with strike zero is equal to the underlying).\\


So in case there is increasing demand for borrowings the AMM should lower the borrowable amounts per collateral unit and, consequently, also lower the strike price at which borrowers can reclaim thei collateral, i.e.,
\begin{equation}
\label{eq:more_borrowing_strike}
\begin{split}
Q_{\mathcal{C}}\uparrow \Rightarrow Q_{\mathcal{B}}\downarrow \Rightarrow K\downarrow.
\end{split}
\end{equation}

And conversely, if there is more lending demand then there will be less collateral available to the AMM, but more borrow currency liquidity, which, in turn, should lead to higher strike prices, i.e.,
\begin{equation}
\label{eq:more_lending_strike}
\begin{split}
Q_{\mathcal{C}}\downarrow \Rightarrow Q_{\mathcal{B}}\uparrow \Rightarrow K\uparrow.
\end{split}
\end{equation}

\subsection{Determining borrowing Costs}
\label{sec:borrowing_costs}
As described in \cref{sec:zero_liquididation_loans_and_vanilla_options}, a user swapping $S$ for $C_K+K$ needs to compensate the AMM for taking on the downside risk. Using the put-call-parity one can see that the swap is fair if
\begin{equation}
\label{eq:put_call_parity}
\begin{split}
\underbrace{C_K+K-P_K}_{\textrm{what borrower get}} = \underbrace{S}_{\textrm{what borrower gives}}
\end{split}
\end{equation}
where $P_K$ denotes the put option with strike price $K$. This means that, in order to make no party worse off, the borrower needs to pay the AMM the put option premium $P_K$. The price of the put option depends on several factors, most importantly on the option's strike price, its time to expiry and the underlying's volatility. However, instead of trying to determine a fair value for the put $P_K$, we will assume the put price $P_K\stackrel{!}{=}X$ is given and let the market infer the corresponding $K$ such that the following equality holds
\begin{equation}
\label{eq:put_call_parity}
\begin{split}
C_K+ \underbrace{K-X}_{\textrm{cash out}} \stackrel{!}{=} S
\end{split}
\end{equation}
, where we refer to $X$ as the \emph{oblivious put price} and $K-X$ is the cash amount the borrower effectively receives. Basically, one could use arbitrary values for $X\in \mathbb R_{\ge 0}$ and always find a corresponding strike $K^*$ that satisfies \cref{eq:put_call_parity}. However, in order to incentive zero-liquidation loans with low LTVs the AMM should steer implied strike prices such that they're lower than the spot value of the collateral, i.e., $K<S$. In other words, $X$ should correspond to the price of an out-of-the-money (OTM) put. One can use a Black-Scholes approximation for an at-the-money put (ATM) as an upper bound value for $X$ which is also easy to compute on-chain, i.e.,
\begin{equation}
\begin{split}
\label{eq:oblivious_put_price}
X = \alpha \cdot  P_{ATM} = \alpha \cdot 0.4 \cdot  S \cdot \sigma \cdot \sqrt{T-t}
\end{split}
\end{equation}
where $\sigma$ denotes the underlying's volatility, $T-t$ is the option's time to maturity and $0<\alpha<1$ is a scaling factor to steer the equilibrium moneyness of the embedded options (or in other words, the LTV of the zero-liquidation loans). Lower $\alpha$ values imply a lower moneyness of the oblivious put, and conversely should steer towards a higher moneyness of the corresponding embedded call option (i.e., lower LTV of the zero-liquidation loan). For example, in the extreme case where $\alpha=0$ then the \emph{oblivious put price} is $X=0$, and the corresponding strike level that satisfies \cref{eq:put_call_parity} is $K=0$ (i.e., maximum moneyness of the call option).\\

So a borrower effectively receives a call option $C_K$ as well as a $K-X$ in the borrow currency. In case the borrower wants to reclaim his collateral at expiry he has to pay back $K$. The maximum payable implied borrowing rate is therefore
\begin{equation}
\begin{split}
R = \frac{X}{K-X}.
\end{split}
\end{equation}



\subsection{Combining Put-Call-Parity with Constant Product AMM}
\label{sec:combining_put_call_parity_with_AMM}
We can now combine the results from \cref{sec:determining_borrowable_amounts} and \cref{sec:borrowing_costs} to derive an AMM for zero-liquidation loans. If we define the strike price to equal $K=\frac{\Delta Q_\mathcal{B}}{\Delta Q_\mathcal{C}}$, it then follows from the put-call-parity\footnote{See \cref{eq:put_call_parity}} that the AMM swaps quantities in the borrow and collateral currency according to
\begin{equation}
\begin{split}
C_{(\Delta Q_{\mathcal{B}}/\Delta Q_\mathcal{C})} + \frac{\Delta Q_{\mathcal{B}}}{\Delta Q_{\mathcal{C}}} - X &= S \\
\Delta Q_\mathcal{C} \cdot C_{K} + \underbrace{\Delta Q_\mathcal{B} - \Delta Q_\mathcal{C} \cdot  X}_{\textrm{borrow currency}} &= \underbrace{\Delta Q_\mathcal{C} \cdot  S}_{\textrm{collateral currency}} \\
\end{split}
\end{equation}
, meaning that when a user pledges $\Delta Q_\mathcal{C} \cdot S$ in the collateral currency (e.g., \verb|ETH|), he then receives $\Delta Q_\mathcal{B} - \Delta Q_\mathcal{C} \cdot X$ in the borrow currency (e.g., \verb|USDC|).

\subsection{Arbitrage}
\label{sec:arbitrage}
Borrowers will have an arbitrage opportunity whenever the implicitly to-be-paid \emph{oblivious put price} $X$ is less than what would be fair according to the put-call-parity, i.e.,
\begin{equation}
\begin{split}
\label{eq:arbitrage_borrowers}
\underbrace{C_{(\Delta Q_\mathcal{B} / \Delta Q_\mathcal{C})} - S}_{\textrm{option's time value}} + \frac{\Delta Q_\mathcal{B}}{\Delta Q_\mathcal{C}} > X \cdot (1+s_{ask})
\end{split}
\end{equation}
where $s_{ask}$ denotes an ask spread parameter, configurable by the AMM. Note that whenever an arbitrage opportunity arises, borrowers will borrow more funds, which according to \cref{eq:more_borrowing_strike} will lead to smaller borrowable amounts per collateral unit and thereby lead to higher implied interest rates, i.e.,
\begin{equation}
\begin{split}
Q_\mathcal{C}\uparrow \Rightarrow Q_\mathcal{B}\downarrow \Rightarrow K\downarrow \Rightarrow R\uparrow.
\end{split}
\end{equation}

Conversely, lenders will have an arbitrage opportunity as soon as the to-be-received \emph{oblivious put price} is larger than what the put-call-parity would imply, i.e.,
\begin{equation}
\label{eq:option_premium_lenders}
\begin{split}
C_{(\Delta Q_\mathcal{B} / \Delta Q_\mathcal{C})} - S + \frac{\Delta Q_\mathcal{B}}{\Delta Q_\mathcal{C}} < X \cdot (1-s_{bid})
\end{split}
\end{equation}
where $s_{bid}$ denotes a bid spread configured at inception of the AMM. If there's such an arbitrage opportunity then the quantity of collateral units will decrease and the amount of liquidity increase, leading to higher borrowing amounts demanded by the AMM per collateral unit, and, ultimately, to lower implied interest rates, i.e.,
\begin{equation}
\begin{split}
Q_\mathcal{C}\downarrow \Rightarrow Q_\mathcal{B}\uparrow \Rightarrow K\uparrow \Rightarrow R\downarrow.
\end{split}
\end{equation}

\subsubsection{Using Flash Loans to arb underpriced Call Options}
\begin{figure}
    \centering
    \includegraphics[width=1.0\textwidth]{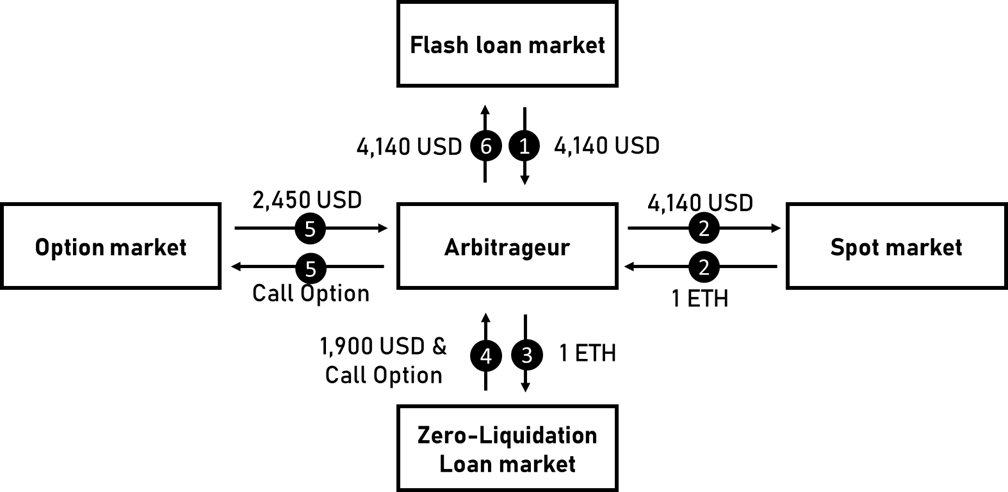}
    \cprotect\caption{\small Using a flash loan to exploit an arbitrage opportunity on the borrowing side.}
    \label{fig:zero_liquidation_loan_use_case}
\end{figure}

Borrowers may use flash loans to instantly take advantage of arbitrage opportunities arising between zero-liquidation loans and option markets. For example, let's assume 1 \verb|ETH| is currently trading at $S=\textrm{4,140 USD}$. Moreover, assume there's currently an offer for a zero-liquidation loan with maturity 31 Dec, that pays out $\textrm{1,900 USD}$\footnote{This upfront amount corresponds to $K-X$ from \cref{eq:put_call_parity}.} upfront for 1 \verb|ETH| collateral for a final repayment amount of 1,920 USD. Lastly, assume there's a call option on \verb|ETH| with expiry 31 Dec and strike $K=\textrm{1,920 USD}$ currently trading for $C_K=\textrm{2,450 USD}$. A user can now do an arbitrage by doing the following steps (see also \cref{fig:zero_liquidation_loan_use_case}):
\begin{enumerate}
\item Borrow 4,140 USD via flash loan 
\item Buy 1 \verb|ETH| for 4,140 USD
\item Pledge 1 \verb|ETH| as collateral in zero-liquidation loan
\item Receive the following two:
\begin{enumerate}
\item 1,900 USD in cash upfront
\item call option to reclaim \verb|ETH| for 1,920 USD
\end{enumerate}
\item Sell call option for 2,450 USD
\item Pay back flash loan and keep arbitrage profit of 210 USD\footnote{This is because: (upfront cash)+(call premium)-(repayment)=1,900+2,450-4,140=210.}
\end{enumerate}
Here, the arbitrage opportunity comes from the fact that the implied call option of the zero-liquidation loan can be sold at a premium, and the upfront cash amount paid out is relatively seen ``too high''. Note that such an arbitrage opportunity assumes that the zero-liquidation loan related option can be transferred and resold in the corresponding open market, i.e., it requires compatibility with the given option market.

\subsubsection{Using 3rd Party Borrowing Markets to arb overpriced Put Options}
\begin{figure}
    \centering
    \includegraphics[width=0.68\textwidth]{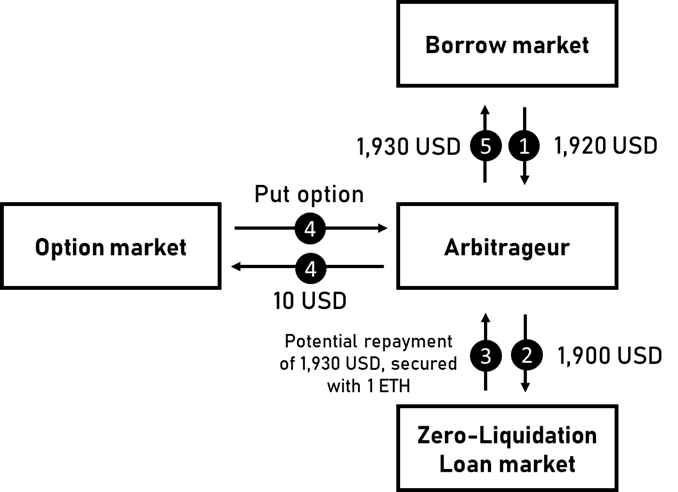}
    \cprotect\caption{\small Steps to exploit an arbitrage opportunity on the lender side.}
    \label{fig:lend_arbitrage_use_case}
\end{figure}

Lenders may take advantage of situations where the oblivious put price applied by the AMM is ``too high'' relative a corresponding option market. For example, let's assume there's a borrowing market that offers a fixed loan of 1,920 USD for a repayment amount of 1,930 USD due at 31 Dec. Further, assume a zero-liquidation loan lets lenders invest 1,900 USD for a possible repayment amount of 1,930 USD at 31 Dec, secured by 1 \verb|ETH| of collateral. Lastly, assume there's an option market where a put option on \verb|ETH| with strike price $K=\textrm{1,930 USD}$ and expiry on 31 Dec is trading for $P_K=\textrm{10 USD}$. A lender can then do an arbitrage as follows (see also \cref{fig:lend_arbitrage_use_case}):
\begin{enumerate}
\item Borrow 1,920 USD with at a fixed repayment amount of 1,930 USD
\item Lend 1,900 USD to the zero-liquidation loan market
\item Secure possible repayment of 1,930 USD, collateralized with 1 
\item Buy put option with strike 1,930 USD
\item Pay back loan at expiry and keep arbitrage profit of 10 USD\footnote{This is because: (repayment from zero-liquidation loan and put option)-(repayment amount due)+(initial borrow amount)-(lent amount)-(put premium paid)=1,930+ 1,920-1,900-10=10.}
\end{enumerate}
The lender is able to make an arbitrage profit because he has effectively hedged his downside risk by buying a put and hence will always be able to repay his original loan of 1,930 USD, whilst having only lent out a ``too small'' amount. In case 1 \verb|ETH| is worth more than the strike price, the lender will receive back 1,930 USD from the zero-liquidation loan market, and otherwise, will get the 1 \verb|ETH| collateral and in addition receive the corresponding put payoff, effectively offsetting his downside risk. At the same time, from the 1,920 USD he borrowed he only had to ``use'' 1,900 USD for the lending transaction and 10 USD to buy the put. Note that a flash loan cannot be used in this scenario because arbing the lending side of a zero-liquidation loan requires cash to be tied up for the corresponding loan duration.

\subsection{Numerical Example}
Let's assume the current \verb|ETH| price is $S_0=4,000$ \verb|USDC|. Further, assume we want to initialize the AMM to provide zero-liquidation loans with an initial $LTV_{init}=83\%$. This can be accomplished by boostrapping the AMM with liquidity contributions $Q_\mathcal{C}=30$ and $Q_\mathcal{B}=100,000$. In this case, the resulting AMM constant is given by $k=3\cdot10^6$.\\

Further, let $\alpha=0.2$, $S_0=4,000$, $\sigma=100\%$ and $\sqrt{T-t}=1.0$ such that the initial \emph{oblivious put price} is $X_{init} = 0.2 \cdot 0.4 \cdot 4,000 \cdot 100\% \cdot 1 = 800$.\footnote{For simplicity, the spread is assumed to be zero.} A borrower could then borrow \verb|2,426 USDC| for \verb|1 ETH|, with a repayment obligation of \verb|3,226 USDC| and an implied maximum APR of $R=24\%$. Conversely, a lender could lend \verb|1,000 USDC| with a possible repayment by the AMM of \verb|1,310 USDC| with a maximum implied APY of $R=24\%$. \cref{tab:example_borrowers} and \cref{tab:example_lenders} provide exemplary terms for borrowing or lending funds to or from the AMM.

\begin{table}[]
\small
\begin{tabular}{p{0.5cm}p{2.2cm}p{2.2cm}p{2cm}p{2cm}p{1.5cm}}
\hline
\textbf{$\Delta Q_\mathcal{C}$} & \textbf{borrower's $CashReceived$} & \textbf{borrower's repayment} & \textbf{implied strike} & \textbf{implied LTV} & \textbf{Max. APR} \\ \hline
1                     & 2,426 USDC                  & 3,226 USDC                    & 3,226 USDC              & 80\%                 & 24\%                    \\ \hline
5                     & 10,286 USDC                   & 14,286 USDC                    & 2,857 USDC              & 71\%                 & 28\%                    \\ \hline
10                    & 17,000 USDC                  & 25,000 USDC                   & 2,500 USDC              & 63\%                 & 32\%                    \\ \hline
20                    & 24,000 USDC                  & 40,000 USDC                   & 2,000 USDC              & 50\%                 & 40\%                    \\ \hline
\end{tabular}
\label{tab:example_borrowers}
\caption{Numerical example of terms for different borrowing amounts $\Delta Q_\mathcal{C}$, assuming $S_0=4,000$.}
\end{table}

\begin{table}[]
\small
\begin{tabular}{p{0.5cm}p{2.2cm}p{2.2cm}p{2cm}p{2cm}p{1.5cm}}
\hline
\textbf{$\Delta Q_\mathcal{C}$} & \textbf{lender's $CashPaid$} & \textbf{AMM's repayment} & \textbf{implied strike} & \textbf{implied LTV} & \textbf{Max. APY} \\ \hline
0.03                     & 100 USDC                   & 132 USDC             & 3,338 USDC                  & 83\%                 & 24\%                    \\ \hline
0.196                     & 500 USDC                  & 657 USDC            & 3,355 USDC              & 84\%                 & 24\%                    \\ \hline
0.388                    & 1,000 USDC                  & 1,310 USDC            & 3,377 USDC              & 84\%                 & 24\%                    \\ \hline
3.382                    & 10,000 USDC                  & 12,706 USDC            & 3,757 UDSC              & 94\%                 & 21\%                     \\ \hline
15.897                    & 100,000 USDC                  & 112,717 USDC            & 7,090 UDSC              & 177\%                 & 11\%                     \\ \hline
\end{tabular}
\caption{Numerical example of terms for different lending amounts $CashPaid$, assuming $S_0=4,000$.}
\label{tab:example_lenders}
\end{table}

\subsection{No-Shortfall Condition}
The AMM shall always be able to exercise the call options it holds or acquires. Thus it needs to be ensured that the AMM's liquidity inventory $Q_\mathcal{B}$ is sufficient to cover possible repayment costs with lenders. This is to prevent shortfall scenarios, where the AMM might end up unable to return collateral to borrowers because it doesn't have enough funds to exercise its own call options with lenders and reclaim collateral. Therefore, it needs to be ensured that the total outstanding possible repayment amounts towards lenders (see \cref{eq:option_premium_lenders}) can never exceed the available funds. More specifically,
\begin{equation}
\begin{split}
\small
\underbrace{ \sum_{i \in L} \Delta Q_{\mathcal{C},i} \cdot X_i \cdot (1-s_{bid})  + \cancel{\Delta Q_{\mathcal{B},i}} }_{\textrm{worst-case amount to be paid by AMM}} < \underbrace{ Q_{\mathcal{B},0} - \sum_{j \in B} \Delta Q_{\mathcal{B},j} + \cancel{\sum_{i \in L} \Delta Q_{\mathcal{B},i}} }_{\textrm{worst-case available liquidity}}
\end{split}
\end{equation}
, where $L$ denotes the list of lenders, $B$ the list of borrowers and $Q_{\mathcal{B}, 0}$ the initially available liquidity at inception of the AMM. Note that the here mentioned \emph{no-shortfall condition} is conservative in that it neglects possible borrower repayments, which would lead to more liquidity being available to the AMM.

\subsection{Simulation}
\begin{figure}
    \centering
    \includegraphics[width=1.0\textwidth]{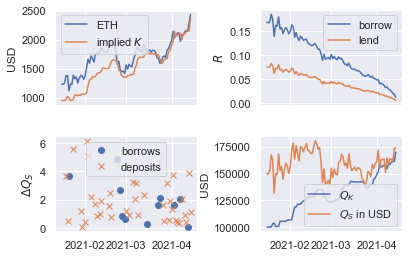}
    \cprotect\caption{\small Backtest results of a hypothetical 90-day zero-liquidation loan market.}
    \label{fig:simulation}
\end{figure}

One can simulate the previously described AMM on historical \verb|ETH/USD| price data. \cref{fig:simulation} illustrates a backtest of a hypothetical 90-day zero-liquidation loan market, where borrowers and lenders trade against the AMM as soon as arbitrage opportunities arise (see \cref{sec:arbitrage}). The plot at the top left shows the price evolution of \verb|ETH| as well as how the implied strike price $K$ changes as borrowers and lenders trade with the AMM. One can see that $K$ tends to stay below the current spot price. This is because in the backtest the parameter for the oblivious put price from \cref{eq:oblivious_put_price} was set to $\alpha=0.5$, implying an ITM option. At expiry, the strike $K$ converges to $S_T$, which is because the oblivious put price and call option go to zero as well, such that \cref{eq:put_call_parity} implies $K=S$.\\

The plot at the top right shows the implied borrow and deposit rates. In the backtest, the spread parameters were set to $s_{bid}=0.5$ and $s_{ask}=0.1$. One can see that the spread between the borrow and deposit rates vanishes over time, which is caused by the fact that the spread is applied on the oblivious put price $X$ (see \cref{sec:arbitrage}), which eventually converges to zero. For the same reason the rates themselves also approach zero at expiry.\\

The plot at the bottom left show the arbitrage trades of borrowers and lenders. One can see that at around \verb|2021-03| the \verb|ETH| price drops from \verb|2,000 USD| to \verb|1,500 USD|, which causes borrowers to start trading against the AMM. This is because the price drop causes the oblivious put price to decrease, i.e., $X\downarrow$, creating arbitrage opportunities for borrowers who, according to \cref{eq:arbitrage_borrowers}, start pushing the strike price downwards to reach an equilibrium state again. Conversely, more lenders tend to trade against the AMM in phases of upward trending \verb|ETH/USD| prices.\\

The plot at the bottom right shows the AMM's inventory in $Q_\mathcal{B}$ and $Q_\mathcal{C}$. One can see that, at inception, the AMM was bootstrapped with an initial $Q_\mathcal{C}$ contribution worth \verb|1.5x| the initial $Q_\mathcal{B}$ liquidity provisions in USD terms. Over time, the $Q_\mathcal{B}$ inventory tends to increase, which is caused by the upward trend in the \verb|ETH/USD| price, that, in turn, creates arbitrage opportunities for lenders, who then start giving liquidity $\Delta Q_\mathcal{B}$ and upside $C_K+K$ to the AMM in return for the oblivious put price premium $X$.\\

\begin{figure}
    \includegraphics[width=.49\textwidth]{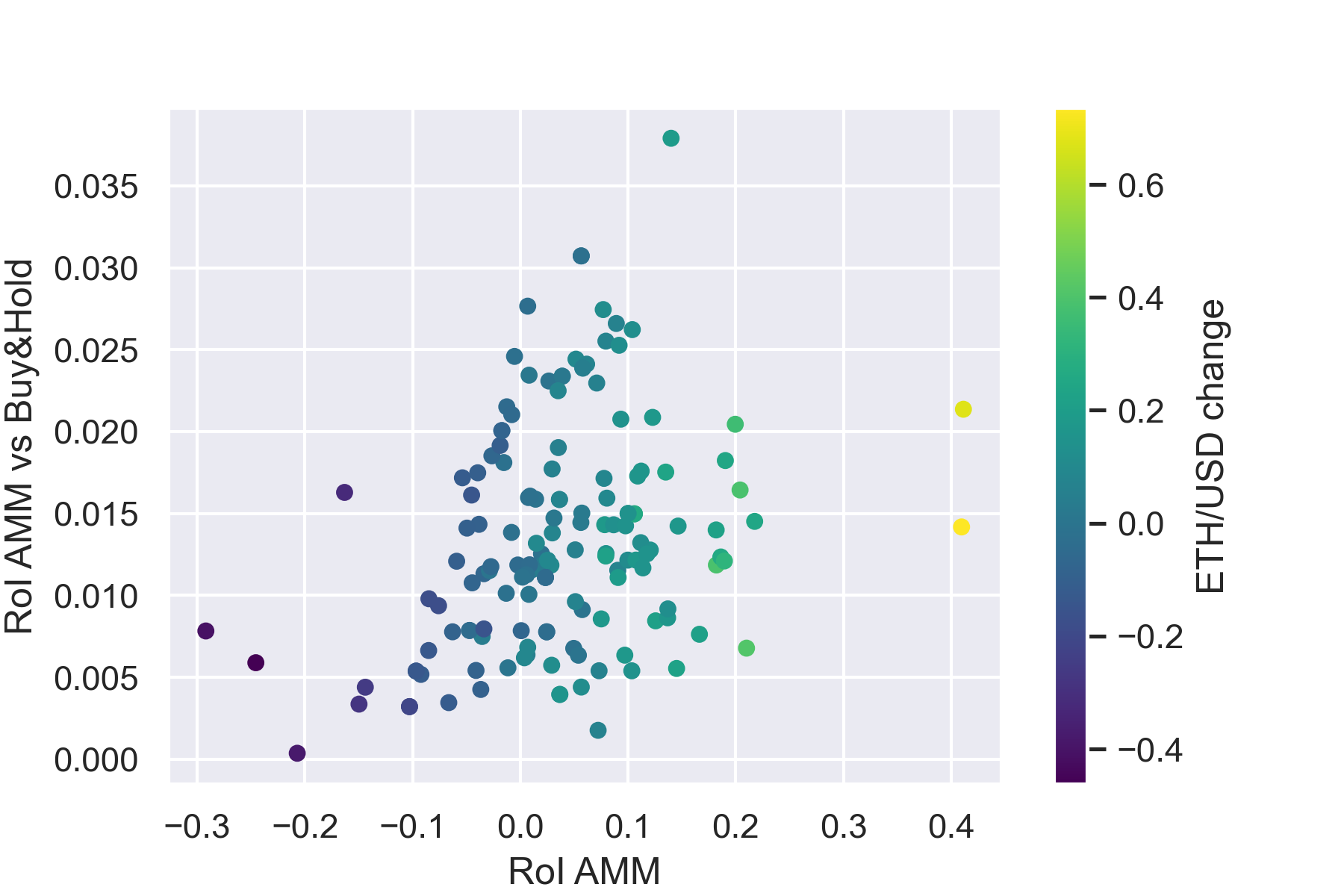}\hfill
    \includegraphics[width=.49\textwidth]{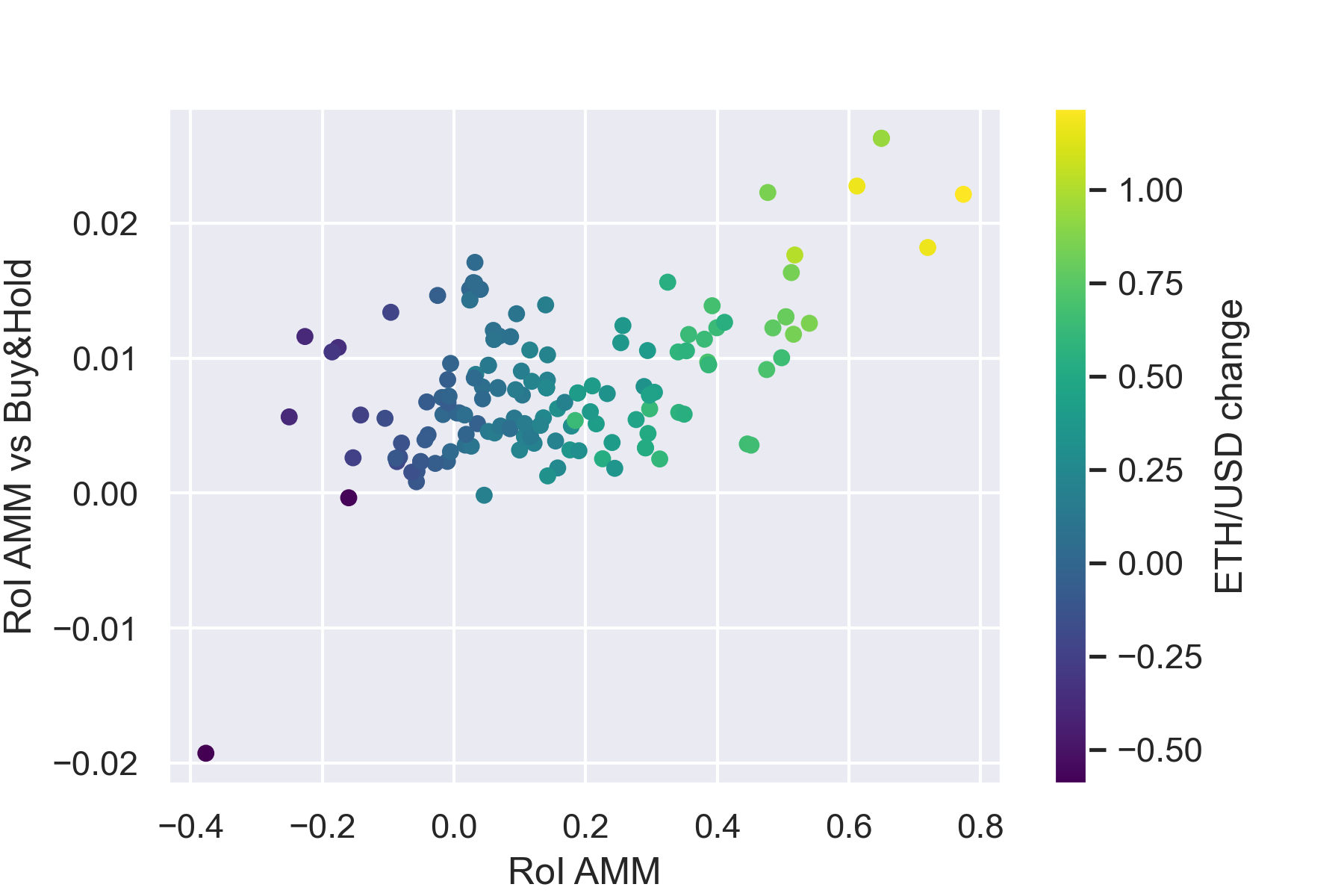}
    \\[\smallskipamount]
    \includegraphics[width=.49\textwidth]{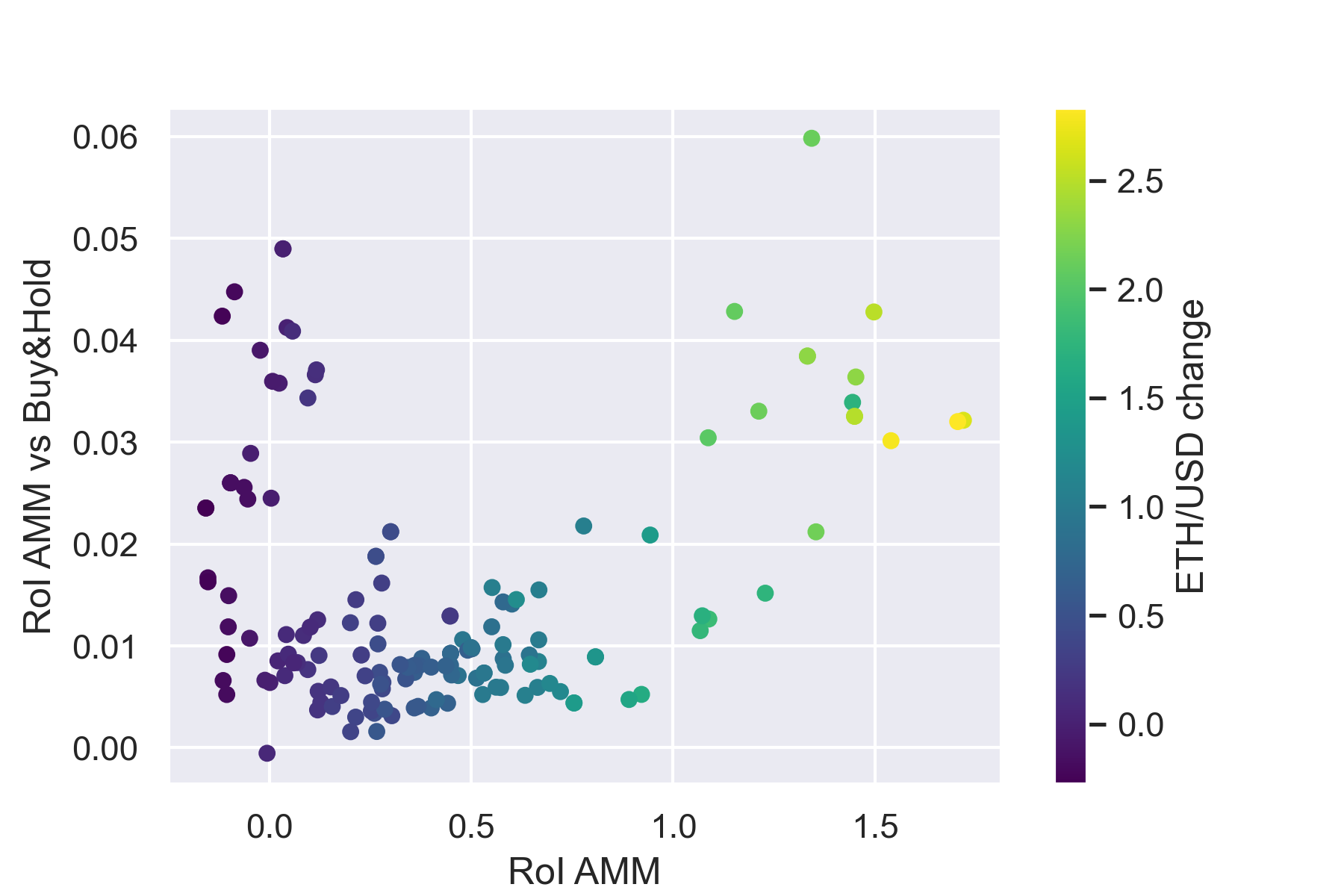}\hfill
    \includegraphics[width=.49\textwidth]{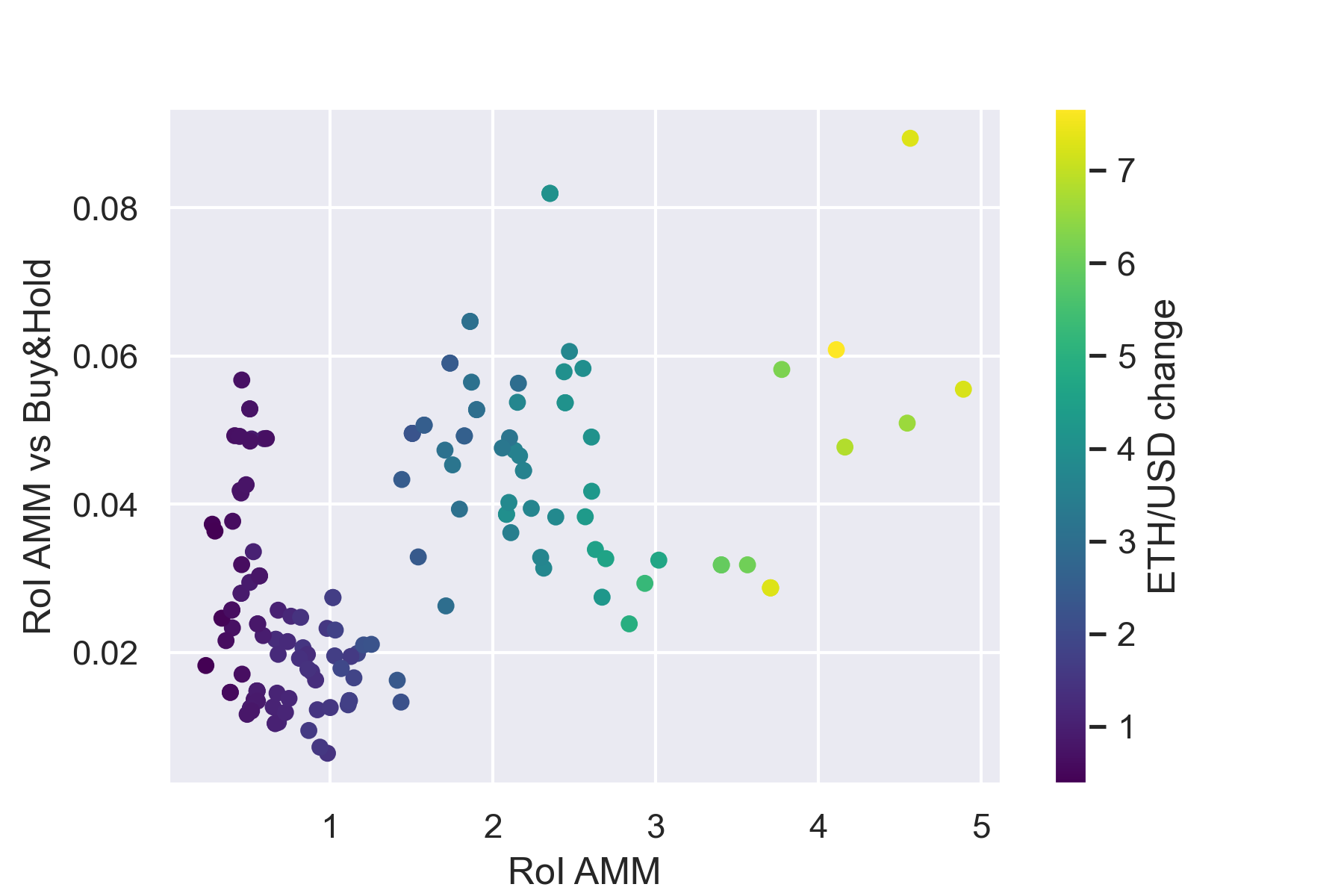}
    \caption{Backtest PnL results for 10 days (top left), 30 days (top right), 90 days (bottom left) and 180 days (bottom right) zero-liquidation loan markets.}\label{fig:backtest_pnl}
\end{figure}

\cref{fig:backtest_pnl} summarizes how the AMM's PnL would have performed under different market scenarios (again using using historical \verb|ETH/USD| price data for backtesting). The plots compare the AMM's RoI with the RoI of a simple buy\&hold strategy, where an investor, instead of investing $Q_\mathcal{B}$ and $Q_\mathcal{C}$ into a zero-liquidation loan AMM would have just held $Q_\mathcal{B}$ and $Q_\mathcal{C}$ during the same time frame. The four plots show backtest results for different time horizons, i.e., the top left is for a 10 days market, the top right for a 30 days, the bottom left for a 90 days and the bottom right for a 180 days market. One can see that the RoI of the AMM tends to outperform a simple buy\&hold strategy by up to 8\%. The relative outperformance tends to be higher for longer dated markets, e.g., up to 3.5\% for 10-day dated and up to 8\% for 180-day dated markets. Moreover, one can observe a positive correlation between the AMM's RoI and the relative outperformance. The AMM's RoI and relative outperformance tends to be highest when \verb|ETH/USD| prices are increasing and, as expected, lowest when prices are falling.



\section{Closing Remarks}
Zero-liquidation loans make DeFi borrowing easier. They eliminate the need to monitor borrowing costs, LTVs, health factors etc., and thereby reduce administrative and operational overhead for borrowers. At the same time, they also offer new yield opportunities for liquidity providers and lenders. By providing an alternative to the otherwise liquidation-centered design approach of current DeFi lending and borrowing protocols, the risk of financial contagion can be reduced and fire sales of collateral assets under stressed market conditions can be avoided. Because zero-liquidation loans are settled without requiring on-chain price data from oracles, they are also more robust against flash crashes. While the herein presented AMM is used to facilitate zero-liquidation loans, its design can be adapted to other option related payoff structures and structured products as well (e.g., reverse convertibles). 

\newpage
\bibliographystyle{abbrv}
\bibliography{main}

\end{document}